# Knowledge Maps and Information Retrieval (KMIR)

Peter Mutschke[1], Andrea Scharnhorst[2], Christophe Guéret[3], Philipp Mayr[4], Preben Hansen[5], Aida Slavic[6]

[1] GESIS – Leibniz Institute for the Social Sciences, Cologne, Germany
[2] Royal Netherlands Academy of Arts and Sciences (KNAW) / Data Archiving and Networked Services (DANS), Amsterdam / The Hague, The Netherlands
[3] Royal Netherlands Academy of Arts and Sciences (KNAW) / Data Archiving and Networked Services (DANS), Amsterdam / The Hague, The Netherlands
[4] GESIS – Leibniz Institute for the Social Sciences, Cologne, Germany
[5] University of Stockholm, Department of Computer and Systems Sciences, Sweden
[6] UDC Consortium, The Hague, The Netherlands

Contact: peter.mutschke@gesis.org

**Abstract:** Information systems usually show as a particular point of failure the vagueness between user search terms and the knowledge orders of the information space in question. Some kind of guided searching therefore becomes more and more important in order to precisely discover information without knowing the right search terms. Knowledge maps of digital library collections are promising navigation tools through knowledge spaces but still far away from being applicable for searching digital libraries. However, there is no continuous knowledge exchange between the "map makers" on the one hand and the Information Retrieval (IR) specialists on the other hand. Thus, there is also a lack of models that properly combine insights of the two strands. The proposed workshop aims at bringing together these two communities: experts in IR reflecting on visual enhanced search interfaces and experts in knowledge mapping reflecting on visualizations of the content of a collection that might also present – visually – a context for a search term. The intention of the workshop is to raise awareness of the potential of interactive knowledge maps for information seeking purposes and to create a common ground for experiments aiming at the incorporation of knowledge maps into IR models at the level of the user interface.

## Motivation, Goals, Objectives, Outcome

The success of an information system depends mainly on its ability to properly support interaction between users and information. Current information systems, however, show as a particular point of failure the vagueness between user search terms and the knowledge orders of the information space in question (Mayr et al. 2008, Mutschke et al. 2011). Studies in interactive information seeking behavior have confirmed that the ability to browse an information space and observe similarities and dissimilarities between information objects is crucial for accidental encountering and the creative use of information (Nicholas et al. 2004, Westerman et al. 2005). This is in particular true for heterogeneous information spaces within the open web. Some kind of guided searching therefore becomes more and more important in order to precisely discover information without knowing the right search terms. Yet, this seems to remain the weakest point of interactive information systems (Ford 2000, Foster 2004, Tang 2007).

Knowledge mapping encompasses all attempts to use visualizations to gain insights into the structure and evolution of large-scale information spaces. Knowledge maps can take the form of network visualizations, treemaps or specific, map like arrangements of search results (cf. Börner et al. 2003, Shiffrin/Börner 2004, Börner 2010, Klavans/Boyack 2010, Skupin et al. 2013, Sahal et al. 2013,

Boyack/Klavans 2013). As an activity performed in very different disciplines – and often independent from each other – it stands in line with the dominance of *the visual* in our culture (Manovich 2009). Knowledge maps of digital library collections are promising navigation tools through knowledge spaces but – to the best of our knowledge – still far away from being applicable for searching digital libraries. Most maps are made for special purposes, are static, and usually not interactive (Akdag Salah et al. 2012). In interactive information systems the use of visual elements to enhance information seeking and discovery is a recurring research issue. However, not much of the experiences made in knowledge mapping have ever been implemented in online interfaces to digital libraries and collections (Börner/Chen 2001), nor is there a stable and continuous knowledge exchange between the "map makers" on the one hand and the Information Retrieval (IR) specialists on the other hand. Thus, there is also a lack of models that properly combine insights of the two strands, which are driven by quite different epistemic perspectives.

Our workshop aims at bringing together these two communities: experts in IR reflecting on visual enhanced search interfaces and experts in knowledge mapping reflecting on visualizations of the content of a collection that might also present – visually – a context for a search term. The intention of the workshop is to raise awareness of the potential of interactive knowledge maps for information seeking purposes and to create a common ground for experiments aiming at the incorporation of knowledge maps into IR models at the level of the user interface. The major focus of the workshop is on the question of how knowledge maps can be utilized for scholarly information seeking in large information spaces. Our interests include interactive IR, information seeking behavior, knowledge mapping, science modelling, information visualization, and digital libraries. The workshop is closely related to the COST action KNOWeSCAPE[1] (Analyzing the dynamics of information and knowledge landscapes) which aims at implementing new navigation and search strategies based on insights of the complex nature of knowledge spaces as well as visualization principles for knowledge maps.

The long-term research goal is to develop and evaluate new approaches for combining knowledge mapping and IR. More specifically, we address questions such as:

- What are appropriate interactive knowledge maps for IR systems?
- How can knowledge maps be utilized for information seeking purposes?
- How to locate an information need on a knowledge map?
- How can (visually enhanced) search interfaces to knowledge maps look like?
- How can interaction with knowledge maps be transformed into IR tasks?
- Can knowledge maps improve searching in large, in particular heterogeneous, cross-language, cross-domain information spaces?
- And the other way around: Can insights from IR also improve knowledge mapping itself?

The availability of new IR test collections that contain citation and bibliographic information like the iSearch collection (see Lykke et al. 2010) or the ACL collection (Ritchie et al. 2006) could deliver an interesting playground for developing or evaluating combined models of IR and knowledge mapping for scholarly searching.

## Topical Outline

To support the previously described goals the workshop topics include (but are not limited to) the following:

- Knowledge maps for digital libraries
- Knowledge orders of information spaces
- Information seeking behaviour
- Information discovery
- Interactive IR systems

---

[1] http://knowescape.org/

- Human–computer IR
- Knowledge Visualization in IR
- Visual interfaces to information systems
- Browsing and navigating information spaces
- Task based user modelling, interaction and personalization
- Evaluation of interactive IR systems.

## Format and Structure

We propose a half-day workshop that initially brings together researchers in IR and knowledge mapping. The workshop will start with an inspirational keynote given by a major researcher in knowledge mapping in order to kick-start thinking and discussion on the workshop topic. This will be followed by a number of paper presentations in the following format: 10 minutes lightning talk for each paper, 20 minutes discussion in groups among the workshop participants followed by 1-minute pitches from each group on the main issues discussed and lessons learned. The workshop will conclude with a round-robin discussion of how to progress in enhancing IR with knowledge mapping approaches.

## Audience

The audiences of IR and knowledge mapping are to a large extent identical: students and scholars as well as science policy makers and the general public seeking for information. Traditional IR serves individual information needs and is – consequently – embedded in libraries, archives and collections alike. Knowledge mapping has grown out from information visualization and only lately been applied as reference system (base maps) for mapping scholarly activity in the context of research policy and evaluation (see Sci2[2]) or as interfaces for web applications (see vivoweb.org). Therefore, the perspectives and approaches are different. Science or knowledge map makers aim for better visualizations. IR is oriented more towards performance characteristics of a search in terms of precision and recall, than to the actual presentation of search results. Consequently, the workshop addresses researchers from IR as well as knowledge mapping having an interest in visually enhanced approaches for (scholarly) searching and information discovery.

## Types of Submissions

Full Papers (6 to 8 pages): Full papers, describing advanced or completed work
Short Papers (4 pages): Position papers or work in progress
Poster and Demonstrations (2 pages): Poster and Presentation of systems or prototypes

Submissions have to follow the Springer LNCS Author Guidelines[3] and should be submitted as PDF files to EasyChair. All submissions will be reviewed by at least two independent reviewers. At least one author per paper needs to register for the workshop and attend the workshop to present the work. In case of no-show the paper (even if accepted) will be deleted from the proceedings AND from the program.

## Output

Printed proceedings will be distributed to all attendees. In addition, workshop proceedings will be deposited online in the CEUR workshop proceedings publication service (ISSN 1613-0073) – This way the proceedings will be permanently available and citable (digital persistent identifiers and long term preservation).

---

[2] https://sci2.cns.iu.edu/user/index.php
[3] http://www.springer.com/computer/lncs?SGWID=0-164-6-793341-0

## Proposed Program Committee (to be confirmed)

Alkim Almila Akdag Salah, University of Amsterdam (The Netherlands)

Nicholas J. Belkin, Rutgers University (USA)

Katy Börner, Indiana University (USA)

Kevin W. Boyack, SciTech Strategies, Inc. (USA)

Robert Capra, University of North Carolina (USA)

Edward A. Fox, Virginia Polytechnic Institute and State University (USA)

Norbert Fuhr, University of Duisburg-Essen (Germany)

Peter Ingwersen, Royal School of Library and Information Science (Denmark)

Claus-Peter Klas, University of Hagen (Germany)

Birger Larsen, Royal School of Library and Information Science (Denmark)

Lev Manovich, City University of New York (USA)

Vivien Petras, Humboldt-Universität zu Berlin (Germany)

André Skupin, San Diego State University (USA)

Catherine L. Smith, Kent State University (USA)

Howard D. White, Drexel University (USA)

Max L. Wilson, University of Nottingham (UK)

## Short bios of the proposers

**Peter Mutschke:** Peter Mutschke is senior researcher at GESIS – Leibniz Institute for the Social Sciences (Cologne) and acting head of the GESIS department "Knowledge Technologies for the Social Sciences". His research focuses on information retrieval, network analysis and Social Web. He worked in a number of national and international research projects such as DAFFODIL (Distributed Agents for User-Friendly Access of Digital Libraries), INFOCONNEX (interdisciplinary information network for Social Sciences, Education Science and Psychology), IRM (value-added search services), the DELOS/NSF Working Group on reference models for digital libraries, and the EU-funded projects WeGov (Where eGovernment meets the eSociety) and SENSE4US (Data insights for policy makers & citizens). Currently, he is involved in major national and European research networks such as the COST Action KNOWeSCAPE ("Analyzing the dynamics of information and knowledge landscapes") and the research network "Science 2.0" of the German Leibniz Association. For both research networks Peter Mutschke is member of the management committee. Peter is author of a number of research articles, member of a number of international programme committees, and was co-organizer of the workshop "Combining Bibliometrics and Information Retrieval" at ISSI 2013 and "Bibliometric-enhanced Information Retrieval (BIR)" at ECIR 2014.

**Andrea Scharnhorst:** Dr. Andrea Scharnhorst is head of e-Research at the Data Archiving and Networked Services (DANS) institution in the Netherlands, a large digital archive for research data primarily from the social sciences and humanities. She also coordinates the computational humanities programme at the e-humanities group of the Royal Netherlands Academy of Arts and Sciences (KNAW) in Amsterdam. Starting in physics (Diploma in Statistical Physics) she got her PhD in philosophy of science. She co-edited books in the Springer Series of Understanding Synergetics on Innovation Networks (with A. Pyka) and recently on Models of Science Dynamics (with K. Boerner and P. van den Besselaar). Her current work in the information sciences is devoted to the development of *knowledge maps* for library collections, research data bases and online knowledge spaces such as Wikipedia. Andrea was co-organizer of the workshop "Combining Bibliometrics and Information Retrieval" at ISSI 2013.

**Christophe Guéret:** Christophe Guéret is a research associate at Data Archiving and Networked Services and the eHumanities group of the Royal Netherlands Academy of Arts and Sciences (KNAW) in Amsterdam. His research activities are around the design of decentralized interconnected knowledge systems and their social implications. Christophe is a co-leader of the Workgroup 4 (Data curation and navigation based on knowledge maps) of the COST Action KNOWeSCAPE ("Analyzing the dynamics of information and knowledge landscapes").

**Philipp Mayr:** Philipp Mayr is a postdoctoral researcher and team leader at the GESIS – Leibniz Institute for the Social Sciences department "Knowledge Technologies for the Social Sciences". Philipp is a graduate of the Berlin School of Library and Information Science at the Humboldt University Berlin where he finished his doctoral research in 2009. Philipp is a member of the European NKOS network and published widely in the areas Informetrics, Information Retrieval and Digital Libraries. He is member of the editorial board of the journals Scientometrics and Information - Wissenschaft & Praxis. His research interests include non-textual ranking in digital libraries, bibliometric methods, evaluation of information systems and knowledge organising sytems, as well as applied informetrics. Philipp was the main organizer of the workshop "Combining Bibliometrics and Information Retrieval" at ISSI 2013.

**Aida Slavic**: Aida Slavic is the editor-in-chief of the Universal Decimal Classification (UDC) and works on the development and the maintenance of this knowledge classification scheme for the UDC Consortium (The Hague). She holds a PhD in library and information studies from the University College London. Her doctoral research was on the use of knowledge classification in the networked environment and her research interest is in the field of knowledge organization, classification, metadata and semantic technologies. Aida is a member of the Scientific Academic Committee of the International Society for Knowledge Organization (ISKO) and member of the editorial board of the Knowledge Organization Journal. She is a co-leader of the Workgroup 1: Phenomenology of Knowledge Spaces of the COST Action KNOWeSCAPE and co-organizer of the first project workshop Knowledge Orders and Science.

**Preben Hansen:** Preben Hansen is Associate Professor at the Department of Computer and Systems Sciences at the University of Stockholm. His research interests address collaborative information seeking and retrieval, task-based and context-based information access, patent IR, evaluation issues and user studies in real-life environments. Preben was involved in a number of EU-projects (Promise, ASSETS, Companions, DELOS NoE and Clarity). A particular focus of his research is on empirical studies of users and use of interactive information access systems.